\documentclass[aps,pra,preprint,showpacs,amsmath,amsfonts,amssymb]{revtex4-1}
\usepackage{amsmath,amsfonts,amssymb,color}
\usepackage{amsthm}
\usepackage{leftidx}
\usepackage{graphicx}
\usepackage{dcolumn}
\usepackage{bm}
\usepackage{epstopdf}
\usepackage{epsfig}

\begin{document}
\title{Discrete time crystals in many-body quantum chaos}
\author{Pekik Nurwantoro}
\email{pekik@ugm.ac.id}
\affiliation{%
	Departemen Fisika, CoECS, FMIPA Universitas Gadjah Mada, Yogyakarta Indonesia 55281
}%
\author{Raditya Weda Bomantara}
\email{phyrwb@nus.edu.sg}
\affiliation{%
	Department of Physics, National University of Singapore, Singapore 117543
}%
\author{Jiangbin Gong}%
\email{phygj@nus.edu.sg}
\affiliation{%
Department of Physics, National University of Singapore, Singapore 117543
}%
\date{\today}


\begin{abstract}
Discrete time crystals (DTCs) are new phases of matter characterized by the presence of an observable evolving with $nT$ periodicity under a $T$-periodic Hamiltonian, where $n>1$ is an integer insensitive to small parameter variations. In particular, DTCs with $n=2$ have been extensively studied in periodically quenched and kicked spin systems in recent years. In this paper, we study the emergence of DTCs from the many-body quantum chaos perspective, using a rather simple model depicting a harmonically driven spin chain.  We advocate to first employ a semiclassical approximation to arrive at a mean-field Hamiltonian and then identify the parameter regime at which DTCs exist, with standard tools borrowed from studies of classical chaos.  Specifically,  we seek symmetric-breaking solutions by examining the stable islands on the Poincar\'{e} surface of section of the mean-field Hamiltonian.  We then turn to the actual many-body quantum system, evaluate the stroboscopic dynamics of the total magnetization in the full quantum limit, and verify the existence of DTCs.  Our effective and straightforward approach indicates that in general DTCs are one natural aspect of many-body quantum chaos with mixed classical phase space structure.   Our approach can also be applied to general time-periodic systems, which is thus promising for finding DTCs with $n>2$ and opening possibilities for exploring DTCs properties beyond their time-translational breaking features.

\end{abstract}


\maketitle

\section{Introduction}

Spontaneous symmetry breaking is a well-known concept in physics which underlies the observation of various phenomena such as ferromagnetism, superconductivity, and the formation of solid crystal structures. Except for the time-translational symmetry, the possibility for a ground state of a certain system to spontaneously break any known symmetries in this universe has been well-studied for years. In 2012, Frank Wilczek put forward a proposal to realize a system exhibiting time-translational symmetry breaking (TTSB), termed ``a time crystal", in the framework of a particle in a ring threaded by a magnetic flux \cite{Wilczek}. By Faraday's Law, the act of switching on the magnetic flux induces an electric field, which drives the localized wave packet in the system to move steadily along the ring, thus breaking the continuous time-translational symmetry of the Hamiltonian. Experimental proposal to observe such a time crystal was then suggested by Ref.~\cite{Li} through a persistent rotation of trapped ions governed by a cylindrical symmetric magnetic potential with fractional fluxes.

It was later pointed out in Ref.~\cite{Bruno} that the moving wave packet arising as a result of turning on the magnetic flux does not correspond to the ground state, and that the real ground state of the Hamiltonian is found to be time independent, which signifies the absence of spontaneous time translational symmetry breaking in the system. Shortly after, a no-go theorem was precisely defined and proved, which forbids the existence of (continuous) time crystals in the ground or equilibrium state of a general Hamiltonian \cite{no, watanabe}.

In the following years, the quest for observing time crystals has shifted towards time-periodic systems exhibiting discrete TTSB, instead of continuous TTSB, thus avoiding the conditions for which the previous no-go theorem \cite{no,watanabe} applies. Such discrete time crystals (DTCs) were first proposed by Ref.~\cite{Sacha} in the framework of cold atom setups. By bouncing ultracold atomic clouds on an oscillating mirror under the influence of gravity and initially preparing the system in its steady (Floquet) state, it was shown that the induction of a small perturbation, such as atomic losses or a measurement of particle positions \cite{Sacha, Sacha2}, collapses the system to a more stable state with higher periodicity from the driving. The emergence of such a state is reminiscent to the formation of Bloch states in spatial crystal structures, which can thus be utilized to extend the theory of various condensed matter phenomena to the time domain, such as Mott-insulating phase \cite{Sacha3}, Anderson localization \cite{Sacha3, Sacha4, Sacha5, Sacha6}, temporal-disorder induced many-body localization \cite{Sacha7}, and interacting systems engineering \cite{Sacha8}.

DTCs have also been independently proposed in periodically driven spin-$1/2$ chain systems by other groups \cite{the1, the4, spin1, spin2}, which have stimulated follow up studies \cite{the2,the3,the5,the6,the7,the8} and several experimental realizations \cite{exp1,exp2,exp3,exp4,exp5}. The original proposal in Ref.~\cite{the1} uses interacting spin-$1/2$ chain with sufficiently large disorders (thus resulting in many-body localization (MBL)), subjected to a global spin flip at the end of each period. Formation of DTCs is then signified by the period doubling in the system's total magnetization, which is insensitive to small variations in the system parameters and persists for a long time (thus suggesting a stable time-translational symmetry broken state). In particular, while MBL was originally thought to be an essential ingredient for creating such DTCs, some studies have demonstrated that formation of DTCs in spin systems without disorder (and thus in the absence of MBL) is possible, where robust period-doubling behavior of the total magnetization occurs before thermalization takes place \cite{the5,the6,the7,the8,exp2,exp3,exp4,exp5}.

The discovery of clean DTCs in spin systems has shed light on the similarity with the formulation of such DTCs in cold atoms, since disorder is not necessary in the latter. In particular, given that DTCs in spin systems are easier to be experimentally verified, whereas studying DTCs in cold atoms allow the natural extension of condensed matter physics to the time domain, such a similarity between the two frameworks is essential for combining these two aspects together, thus allowing future experiments to access other promising features of DTCs beyond their TTSB signatures, such as the realization of condensed matter phenomena in the time domain \cite{Sacha3,Sacha4,Sacha5,Sacha6,Sacha7,Sacha8} and quantum computing applications \cite{RG}.  To further establish another similarity between the two frameworks, we explore the formation of DTCs in spin systems under a harmonic driving of period $T$, which is typically used in the cold atoms formulation of DTCs \cite{Sacha}.

As a second and perhaps more important motivation of this work, we advocate to use the language of classical chaos to facilitate our search for
DTCs. Indeed, for a periodically driven one-dimensional classical system with partially regular and partially chaotic phase space structure,  chains of stable islands are representative features of the so-called Poincar\'{e} surface of section. These chains of stable islands naturally represent discrete TTSB solutions insofar as trajectories launched from one of the islands will exhibit periodic hopping among them, with the hopping period equal to a multiple of that of the driving period.  Specifically, to arrive at an effective one-dimensional mean-field Hamiltonian, we employ a semiclassical treatment of a continuously driven spin chain using SU(2) coherent states,  a rather standard semiclassical tool in the literature of quantum chaos.  Our mean-field approach enables us to locate parameter regimes at which period-doubling of an appropriately constructed operator is observed.  Once such period-doubling parameter regimes are identified, we then verify if DTC signatures, namely, the robust long-lasting $2T$-periodicity of the total magnetization, indeed present in the full quantum system.  The full many-body quantum dynamics simulation is carried out by the use of time-dependent density matrix renormalization algorithm.   As such,  this work is really about the dynamics of many-body quantum chaos with a focus on DTC signatures.

This article is structured as follows. In Sec. II we introduce a harmonically driven spin-chain model and set up some notations. In Sec. III we derive its approximate single-particle Hamiltonian in the classical limit, and plot its Poincar\'{e} surface of section \cite{PSOS} to identify the location of its period-doubling islands at several different system parameters. By appropriately initializing the state based on the location of these islands, we verify through the use of density matrix renormalization algorithm that the total magnetization exhibits a robust $2T$ periodicity in the full quantum regime. In Sec. V, we conclude this paper and present some future directions.

\section{Model}

Consider a harmonically driven spin-chain model described by the time-dependent Hamiltonian
\begin{equation}
\label{eqn:1}
\hat{H}(t)=-h\sum_{i=1}^N \hat{\sigma}_i^x\,\cos^2\left(\frac{\omega t}{2}\right)-J\sum_{i=1}^{N-1}\hat{\sigma}_i^z\,\hat{\sigma}_{i+1}^z+\lambda \hat{V}(\hat{\mathbf{\sigma}}),
\end{equation}
where $\hat{\sigma}_i^\kappa \,(\kappa=x,y,z)$ are the Pauli matrices describing spin degree of freedom at site $i$, $h$ is the strength of an external time-periodic magnetic field along the $x$ direction, $\omega$ is the frequency of the drive, $J$ is the nearest neighbor spin-spin interactions, $\hat{V}(\hat{\mathbf{\sigma}})=\sum_{i=1}^N (\sigma_i^y+\sigma_i^z)$ is an additional static magnetic field with $\lambda$ controlling its strength and $N$ is the number of spin sites. If $\lambda=0$, Eq.~(\ref{eqn:1}) describes a periodically driven Ising spin chain, which has been proposed to be a candidate for exploring DTC phases \cite{the3, the5, the6, the8}. In all these previous works \cite{the3, the5, the6, the8}, however, a quenched or Dirac-delta type of periodic driving is usually employed to observe its DTC signatures. Despite its simplicity, such a discontinuous driving is not very natural, especially if one hopes to utilize DTCs for some real life applications in the future. This is what motivates us to instead employ a continuous time dependence of the external magnetic field in our model, and in the following we will present a recipe for identifying its DTC phases.


For a given initial state $|\psi(t_0)\rangle$, the state of system dynamics $|\psi(t)\rangle = \hat{U}(t;t_0) |\psi(t_0)\rangle $ is dictated by the time evolution operator from $t_0$ to $t$ as (take $\hbar=1$)
\begin{equation}
\label{eqn:2}
\hat{U}(t;t_0)=\mathcal{T}\exp\left[-i\int_{t_0}^t\hat{H}(t')\,dt'\right],
\end{equation}
where $\mathcal{T}$ is the time-ordering operator. Since $\hat{H}(t)$ and $\hat{H}(t')$ with $t\neq t'$ do not generally commute with each other, a closed form to Eq.~(\ref{eqn:2}) is not readily available. An exception to this happens at special parameter values $J =\hat{V}(\hat{\mathbf{\sigma}})= 0$, in which case only Pauli matrices $\hat{\sigma}_i^x$ appear in $\hat{H}(t)$. In this noninteracting limit, its time evolution operator over a single period $T=\frac{2\pi}{\omega}$ can be calculated by direct integration,
\begin{eqnarray}
\hat{U}(T;0)&=& \exp\left[-\mathrm{i} \sum_{j=1}^N \int_0^T h\cos^2\left(\frac{\omega t'}{2}\right)\hat{\sigma}_j^{x} dt'\right]\;,\nonumber\\
&=&\exp\left[-\sum_{j=1}^N \mathrm{i} \frac{hT}{2}\hat{\sigma}_j^x\right]\;. \label{flo}
\end{eqnarray}
In particular, by further setting $hT/2=\frac{\pi}{2}$, two period-doubling states can be immediately constructed as $|\psi_\pm (\phi) \rangle = |\pm \pm\cdots \pm \rangle_n$,
which satisfies $\sigma_j^n |\psi_\pm \rangle_n = \pm |\psi_\pm \rangle_n$ with $\sigma_j^n=\cos\phi \sigma_y+\sin\phi \sigma_z$ and $\phi$ is any phase factor.


We now define $\hat{M}=\frac{1}{N} \sum_{j=1}^N \hat{n}\cdot\hat{\sigma}_j$ as the total magnetization in the $\hat{n}=\cos\phi \hat{y}+\sin \phi \hat{z}$ direction. It is easy to verify that $\langle \hat{M} \rangle$, where $\langle\cdots \rangle$ is taken with respect to the time evolution of $|\psi_\pm (\phi)\rangle$, exhibits $2T$ periodicity in the noninteracting case for any $\phi$. At nonzero interaction, such $2T$ periodicity will in general be lost, except possibly for a small range of values of $\phi$. At these $\phi$ values, the $2T$ periodicity of $\langle \hat{M} \rangle$ is insensitive to small variations in other parameter values $h$, $J$, and $\lambda$ and is persisting indefinitely in the thermodynamic limit, thus establishing the DTC phase of our model \cite{the5, the6}. At this point, our problem reduces to identifying this range of $\phi$ values, or if it exists to begin with.

The presence of many unknown variables makes it computationally difficult to tackle this problem directly in the full quantum picture. Instead, we propose to first employ a semiclassical approximation to obtain an effective nonlinear single-particle Hamilton equations of motion.  We then examine its Poincar\'{e} surfaces of section \cite{PSOS}, especially period-doubling stable islands that are sufficiently large in the phase space.  The range of $\phi$ values that gives rise to a $2T$-periodic $\langle \hat{M} \rangle$ then corresponds to the phase space locations within a period-doubling island, which we found to be located in the vicinity of $\phi=0$. Finally, we hope to verify that the obtained parameter regime indeed shows the anticipated DTC signatures in the full quantum regime.

\section{Formation of discrete time crystals}


\subsection{Semiclassical approach}
We first note that in the ideal case of zero interaction and imperfections, a period-doubling state $|\psi_\pm (\phi)\rangle$ defined above can be written in the spinor basis as a tensor product
\begin{eqnarray}
|\psi_\pm (\phi)\rangle &=& \bigotimes_{j=1}^N \left( \begin{array}{c}
\psi_{1,j} \\
\psi_{2,j}
\end{array}\; \right)\;, \label{tpstate}
\end{eqnarray}
where $\psi_{1,j}=\sqrt{\frac{1\pm \sin \phi }{2}}$ and $\psi_{2,j}=\pm \sqrt{\frac{1\mp \sin \phi}{2}}$ for all $j$. In the following, we will consider a class of tensor product states defined in Eq.~(\ref{tpstate}) with general values of $\psi_{1,j}=\psi_1$ and $\psi_{2,j}=\psi_2$, for which the Hamiltonian expectation value becomes

\begin{equation}
\langle H \rangle = -N h\left(\psi_1^*\psi_2+\psi_2^*\psi_1\right)\cos^2\left(\frac{\omega t}{2}\right)-2(N-1)J\left(|\psi_1|^4+|\psi_2|^4\right)+\lambda \hat{V}(\psi_1,\psi_2,\psi_1^*,\psi_2^*)\;
\end{equation}
up to a term independent of $\psi_1$ and $\psi_2$, where we have also employed the identity $1-2|\psi_1|^2|\psi_2|^2=|\psi_1|^4+|\psi_2|^4$ as a consequence of the normalization.

\begin{figure}
	
	\begin{center}
		\includegraphics[scale=0.6]{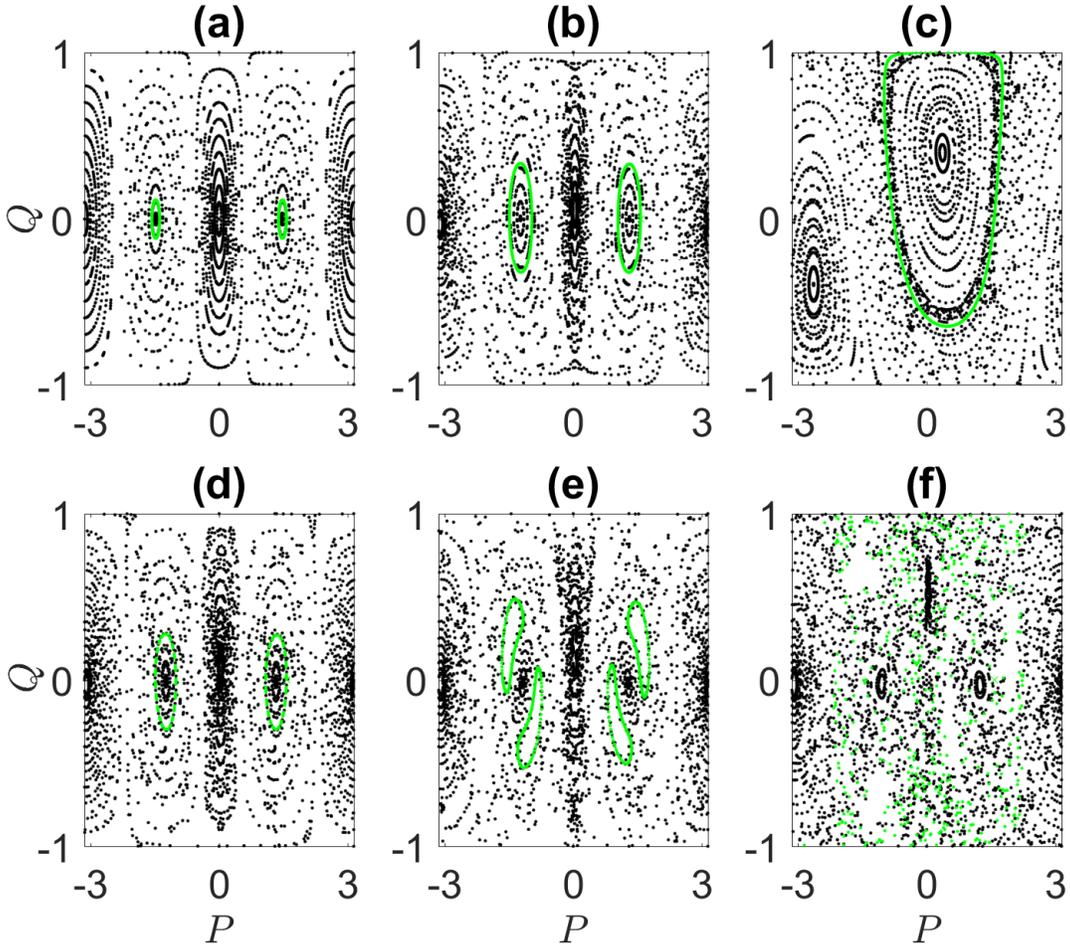}
	\end{center}
	\caption{PSOS governed by Eq.~(\ref{ham}) at different imperfection parameters, i.e., $\epsilon$ and $\lambda$ (panel (a) to (c)), and interaction strength $J$ (panel (d) to (f)). Green marks denote the stroboscopic evolution of a state initially prepared in $(P_0,Q_0)=(\pi/2,0)$ System parameters are chosen as (a) $J T=1$, $\lambda T=\epsilon T=0$, (b) $J T=1$, $\lambda T=\epsilon T=0.05$, (c) $J T=1$, $\lambda T=\epsilon T=0.5$, (d) $J T=2$, $\lambda T=\epsilon T=0.05$, (e) $J T=3$, $\lambda T=\epsilon T=0.05$, and (f) $J T=4$, $\lambda T=\epsilon T=0.05$.}
	\label{sos}
\end{figure}

In the semiclassical limit $N\rightarrow \infty$, the system dynamics is governed by the Hamilton function $\mathcal{H}=\frac{\langle H\rangle}{N}$, which can be recast in terms of canonical variables $Q = \left|\psi_1\right|^2 - \left|\psi_2\right|^2$ and $P = \xi_2 - \xi_1$, where $\xi_1 (\xi_2)$ is the phase of $\psi_1 (\psi_2)$ as
\begin{equation}
\mathcal{H}=-h\sqrt{1-Q^2}\cos(P)\cos^2\left(\frac{\omega t}{2}\right)-J\left(1+Q^2\right)+\lambda \hat{V}(Q,P)\;. \label{ham}
\end{equation}
Through plotting its Poincar\'{e} surface of section (PSOS), which is obtained by solving the Hamilton equations of motion for $Q$ and $P$ and recording the phase space coordinates at each integer multiple of period $T$ \cite{PSOS}, we may probe the DTC signatures of our model by first considering the locations of the period-doubling islands. As depicted in Fig.~\ref{sos}, such period doubling islands exist in the vicinity of $(P,Q)=(\pm\pi/2,0)$, even in the presence of small deviation $\epsilon$ for which $hT/2=\pi/2+\epsilon$ and nonzero $\lambda$. It should be also highlighted that period-doubling islands on PSOS are highly typical in classical systems with a mixed phase space structure and in this sense, discrete TTSB in the classical domain is somewhat ubiquitous.

\begin{figure}
	
	\begin{center}
		\includegraphics[scale=0.6]{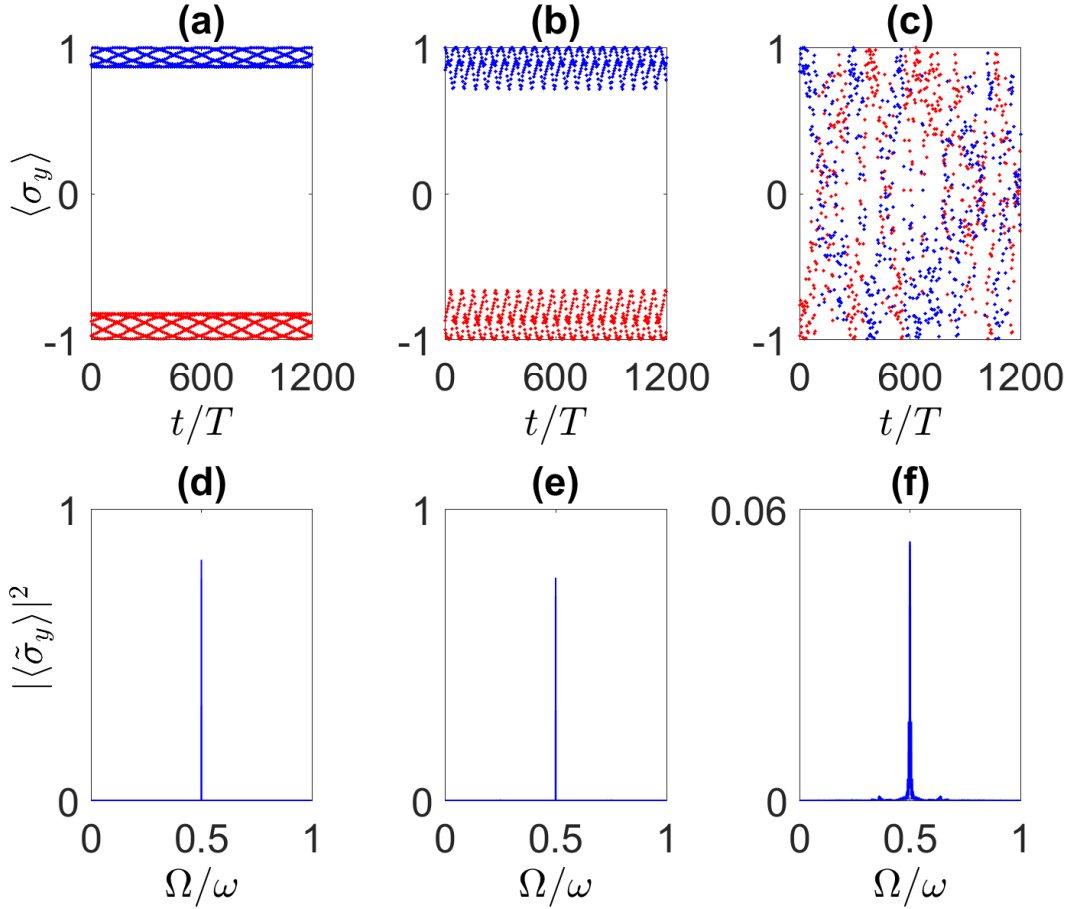}
	\end{center}
	\caption{(a)-(c) Stroboscopic time evolution of $\langle \sigma_y\rangle$ at $\epsilon T=\lambda T=0.05$, $\psi(t=0)=\psi_+$, and different interaction strengths (a) $JT=1$, (b) $JT=3$, and $JT=4$ in the semiclassical setting. (d)-(f) Power spectrum associated with panel (a)-(c) respectively.}
	\label{DTCsc}
\end{figure}

Next, we define a ``semiclassical" spinor state as $\psi=\left(\begin{array}{c}\psi_1 \\ \psi_2 
\end{array}\right)$, such that the phase space coordinate $(P,Q)=(\pm \pi/2,0)$ corresponds to the state

\begin{equation}
\psi_\pm =\frac{1}{\sqrt{2}}\left( \begin{array}{c}
1 \\
e^{\pm\mathrm{i}  \pi/2}
\end{array}\right) \;,
\end{equation}
which are eigenstates of the Pauli matrix $\sigma_y=\left(\begin{array}{cc}
0 & -\mathrm{i} \\
\mathrm{i} & 0 
\end{array}\right)$. Still within the semiclassical limit, we can then define an ``observable" $\langle \sigma_y \rangle(t)=\psi^\dagger(t) \sigma_y \psi(t)=\mathrm{i} \left(\psi_2^*\psi_1 - \psi_2^*\psi_1 \right)$ to represent a quantity that resembles the total magnetization in the $y$-direction, the latter of which is used to probe DTC signatures in the full quantum setting. In Figs.~\ref{DTCsc}(a)-(c), we present the stroboscopic evolution of such an observable at nonzero imperfections up to $1200$ periods for different interaction strengths, where $\psi(t=0)=\psi_+$ and blue (red) marks show the value of $\langle \sigma_y \rangle(t)$ at even (odd) multiples of the period. There, we find that $\langle \sigma_y \rangle$(t) exhibits $2T$ periodicity at moderate interaction strengths and becomes chaotic at stronger interaction strengths, which agrees with the PSOS structures shown in Figs.~\ref{sos}(b), (e), and (f) respectively. To further demonstrate its $2T$ periodicity, we also plot in Figs.~\ref{DTCsc}(d)-(f) the associated power spectrum $\langle \tilde{\sigma}_y\rangle(\Omega)=\frac{1}{\mathcal{N}}\sum_{n=1}^{\mathcal{N}} \langle \sigma_y\rangle (t) e^{\mathrm{i} n \Omega T}$, where $\mathcal{N}$ is the number period and persisting $2T$ periodicity is signified by the existence of a single sharp peak at $\Omega=\frac{\omega}{2\pi}$.

\begin{figure}
	
	\begin{center}
		\includegraphics[scale=0.6]{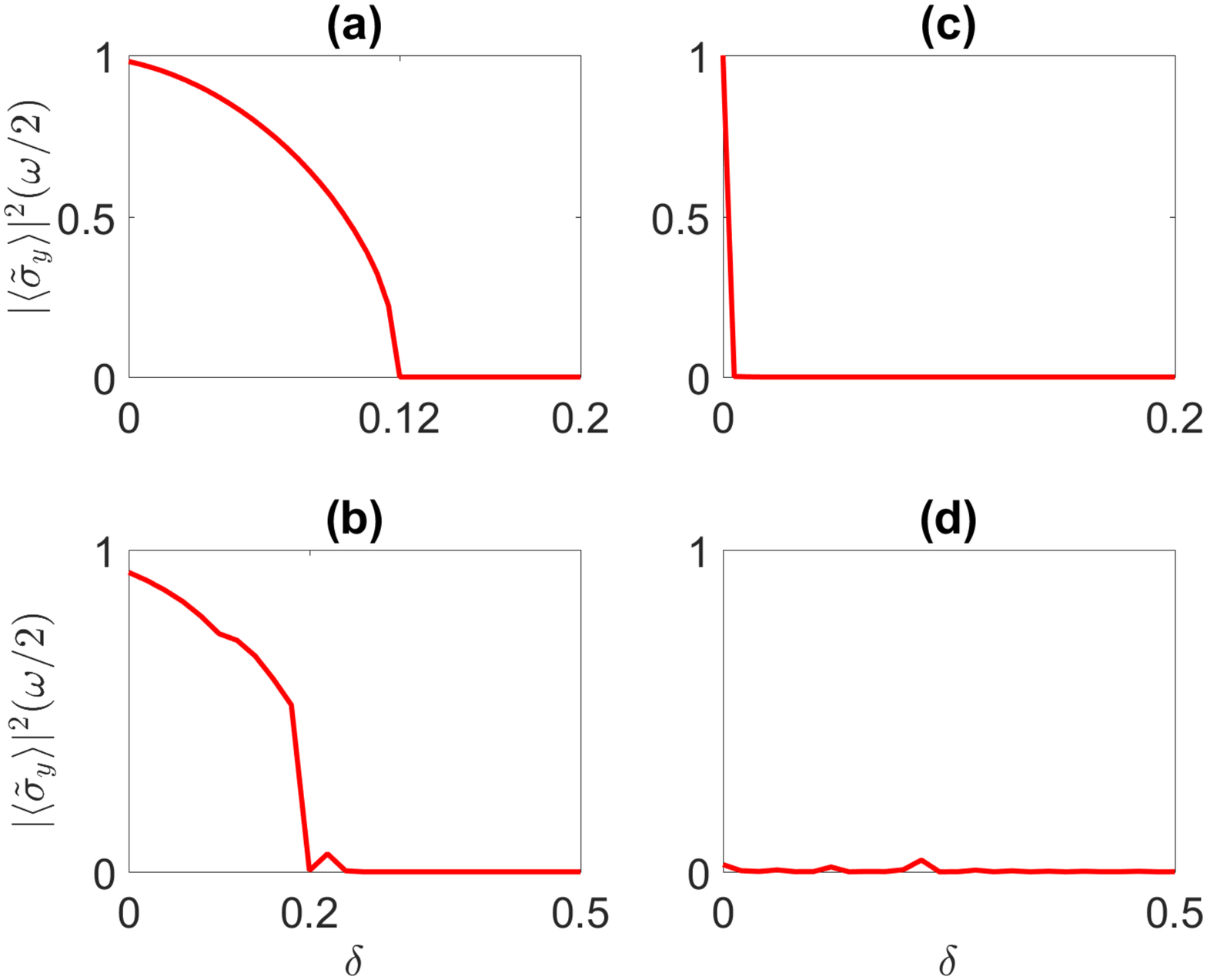}
	\end{center}
	\caption{Subharmonic peak ($|\langle \tilde{\sigma}_y\rangle|^2(\omega/2)$) as $\epsilon T=\lambda T=\delta$ is varied. (a) $JT=1$, (b) $JT=2$, (c) $JT=0$, and (d) $JT=5$.}
	\label{op}
\end{figure}

Finally, in order to identify the DTC phase transition, we plot $|\langle \tilde{\sigma}_y\rangle|^2(\omega/2)$ as the imperfection parameter $\delta=\epsilon=\lambda$ is varied in Fig.~\ref{op}. We find that at moderate interaction strengths, there exists a finite range of values for which $|\langle \tilde{\sigma}_y\rangle|^2(\omega/2)$ exhibits a sharp peak, thus signifying a DTC phase. Moreover, increasing interaction strength in general enlarges the regime in which our system is in the DTC phase (see panel (a) and (b) of Fig.~\ref{op} for comparison), while at the same time also increases the beating structure in the $\langle \sigma_y \rangle (t)$ profile, which causes an overall reduction in the sharpness of $|\langle \tilde{\sigma}_y\rangle|^2(\omega/2)$. As the system enters a chaotic regime at large interaction strengths, the peak at $\Omega=\omega/2$ becomes too small to resolve (Fig.~\ref{op}(f)), leading to the absence of DTC altogether. On the other hand, without any interaction, the peak of $|\langle \tilde{\sigma}_y\rangle|^2(\Omega)$ becomes extremely sensitive to any arbitrarily small variations in $\delta$ (Fig.~\ref{op}(c)), which thus also implies the absence of DTC.

\subsection{Full quantum approach}

The semiclassical results obtained above allow us to probe the parameter regime in which DTC phase is expected to emerge in the full quantum regime, which we will now verify by calculating the stroboscopic dynamics of the total magnetization in the $y$-direction. To this end, we employ time-dependent density matrix renormalization group (t-DMRG) to numerically solve the many-body Schr\"{o}dinger equation for intermediate site number $N$ \cite{feiguin, white1, white2, white3,white4, scholl1, scholl2, scholl3, hallberg, cazalilla1, cazalilla2, luo, vidal1, vidal2}. 

Such a t-DMRG algorithm can be summarized as follows. First, we write the initial state in the matrix product state form. Next, we perform a third-order Suzuki-Trotter expansion \cite{suzuki1,suzuki2,suzuki3} to separate the one-period time-evolution operator into products of time-evolution operators at small time interval $\Delta t$ associated with terms in the Hamiltonian acting on either even or odd lattice sites only. This in turns allows us to write such a one-period time-evolution operator in the matrix product operator form. Applying such an operator to the initial state then amounts to the modification of the matrices within the matrix product state formalism, thus creating a new time-evolved state. The process can then be repeated to further evolve our state by one-period. In general, however, the bond dimension of such a state (the maximum dimension of the matrices generating such a state) increases with each application of the one-period time evolution operator, leading to the increase in complexity as the state evolves at longer times. To reduce the computational time required to execute such an algorithm, a truncation of the bond dimension to a maximum value of $M$ is made. Ideally, higher accuracy is achieved by increasing $M$ and/or decreasing $\Delta t$, which however comes at the cost of longer computational time. It is thus necessary to appropriately choose the values of $M$ and $\Delta t$ to ensure that our results are obtained with a sufficiently good accuracy and completed within a reasonable computational time. To this end, we first observe that $M$ and $\Delta t$ have different effects on the stroboscopic dynamics of the total magnetization in the $y$-direction. That is, variations in $M$ generally affect the decay rate of the total magnetization at early time, whereas variations in $\Delta t$ affect its late time decay. Based on these characteristics, we choose an optimum value for $M$ ($\Delta t$) by making sure that some variations in such a value have insignificant effects on the initial (late) time decay rate. 

\begin{figure}
	
	\begin{center}
		\includegraphics[scale=1.0]{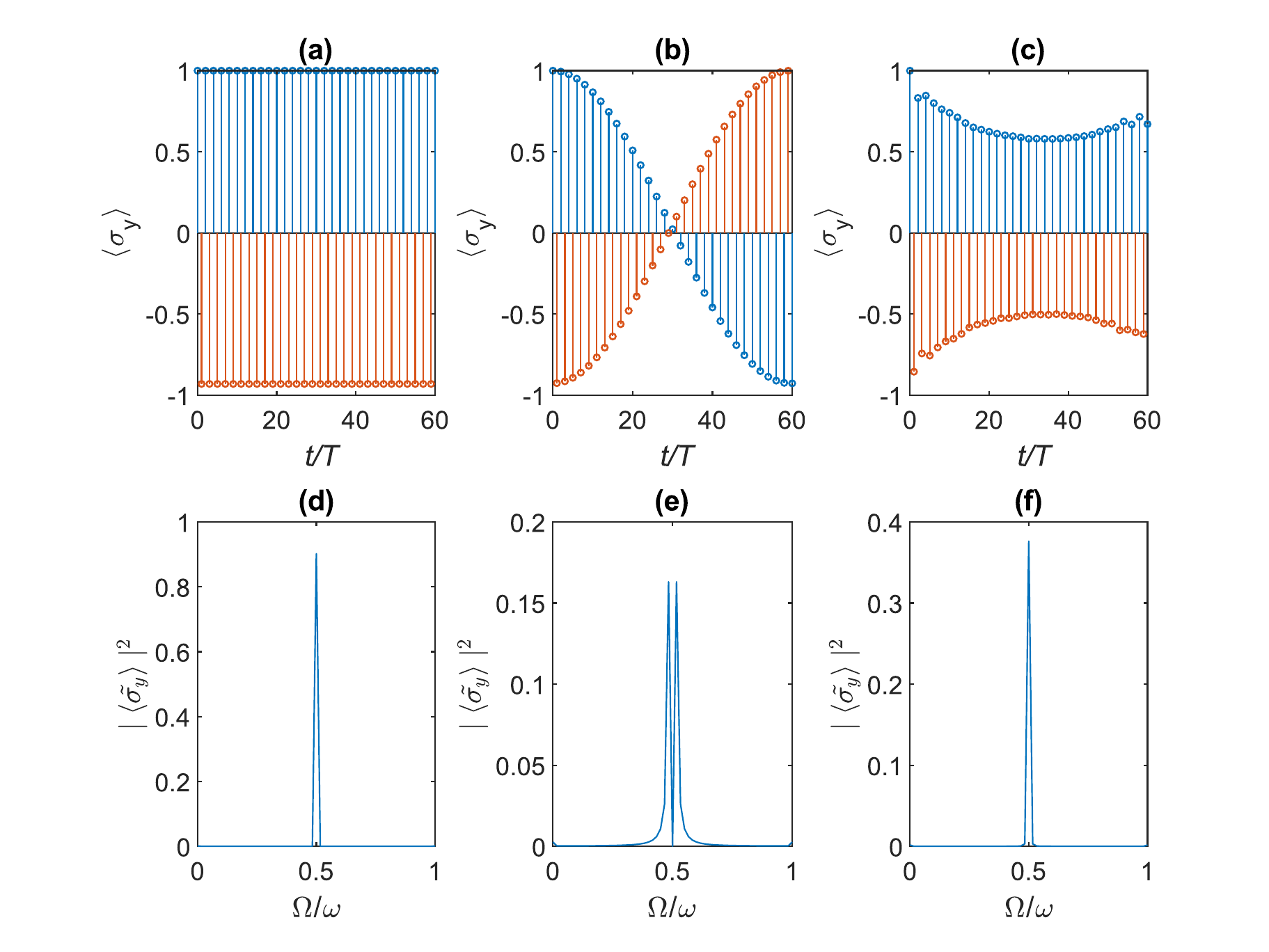}
	\end{center}
	\caption{DMRG results for site number $N=30$, $M=30$, $hT/2=\pi/2$ and $\Delta t/T=0.001$. Stroboscopic time evolution of $\langle\sigma_y\rangle$ and its corresponding power spectrum $\mid\langle\tilde{\sigma_y}\rangle\mid^2$ at different parameters (a) \& (d) $JT=0.0, \epsilon T=0.0, \lambda T=0.0$, (b) \& (e) $JT=0.0, \epsilon T=0.05, \lambda T=0.05$, and  (c) \& (f) $JT=0.5, \epsilon T=0.05, \lambda T=0.05$. In Figs. (a)-(c), blue (red) dots represent the values of $\langle\sigma_y\rangle$ at even (odd) multiples of the period, whereas vertical lines are to guide the eye at their respective dots.}
	\label{fq1}
\end{figure}

Figure~\ref{fq1} shows the time-evolved total magnetization in the $y$-direction ($\langle \hat{M}_y \rangle$) at several interaction strengths and their corresponding power spectrum, with all spins initially aligned in the positive $y$-direction. In particular, we observe from Fig.~\ref{fq1}(b) and (c) that finite interaction is indeed necessary to maintain its $2T$ periodicity in the presence of small perturbations, which agrees with our semiclassical prediction. That is, while $\langle \hat{M}_y \rangle$ shows a perfect $2T$ periodicity in the noninteracting limit at exactly $hT/2=\pi/2$, a small imperfection can already destroy this periodicity, which results in the splitting of the central peak in its power spectrum as shown in Fig.~\ref{fq1}(b). By contrast, at moderate interaction strength, such a $2T$ periodicity remains even in the presence of the same imperfection (see Fig.~\ref{fq1}(c)). 

\begin{figure}
	
	\begin{center}
		\includegraphics[scale=1.0]{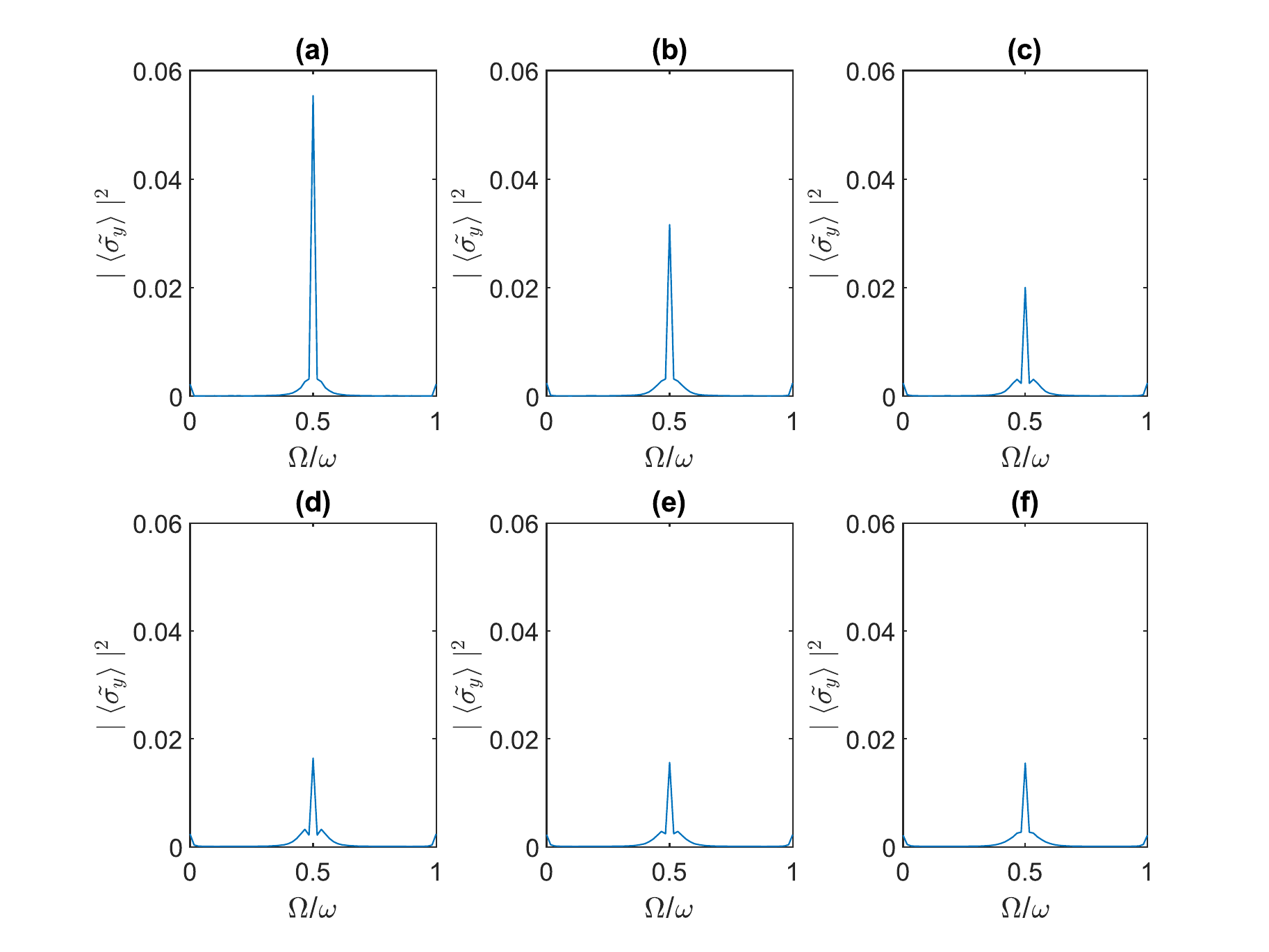}
	\end{center}
	\caption{DMRG results for site number $N=30$, $M=30$, $hT/2=\pi/2$, $\Delta t/T=0.001$. Power spectrum $\mid\langle\tilde{\sigma_y}\rangle\mid^2$ at a fixed interaction strength of $J T=1.0$ for (a) $\epsilon T=0.05, \lambda T=0.05$, (b) $\epsilon T=0.09, \lambda T=0.09$, (c) $\epsilon T=0.13, \lambda T=0.13$, (d) $\epsilon T=0.15, \lambda T=0.15$, (e) $\epsilon T=0.17, \lambda T=0.17$, and (f) $\epsilon T=0.19, \lambda T=0.19$.}
	\label{fq2}
\end{figure}

To verify the DTC phase boundary obtained previously in the semiclassical setting through Fig.~\ref{op}, we also plot $\mid\langle\tilde{\sigma_y}\rangle\mid^2$ at a fixed interaction strength of $J T=1.0$ and various imperfection strengths ($\epsilon=\lambda=\delta$) in Fig.~\ref{fq2}. While $\mid\langle\tilde{\sigma_y}\rangle\mid^2(\omega/2)$ observed there is generally smaller than that observed in Fig.~\ref{op} due to finite-size effect, it shows a similar dependence on $\delta$ as that predicted in the semiclassical setting. In particular, $\mid\langle\tilde{\sigma_y}\rangle\mid^2$ initially decays as $\delta$ increases. At $\delta =0.13$, i.e., beyond the phase boundary observed in Fig.~\ref{op}(a), two additional peaks emerge in the vicinity of $\Omega=\omega/2$, thus signifying the absence of DTC. As $\delta$ further increases, $\mid\langle\tilde{\sigma_y}\rangle\mid^2(\omega/2)$ remains almost constant at a very small value, while the additional peaks tend to spread over all values of $\Omega$, thus removing any periodic structure in the total magnetization.

\begin{figure}
	
	\begin{center}
		\includegraphics[scale=1.0]{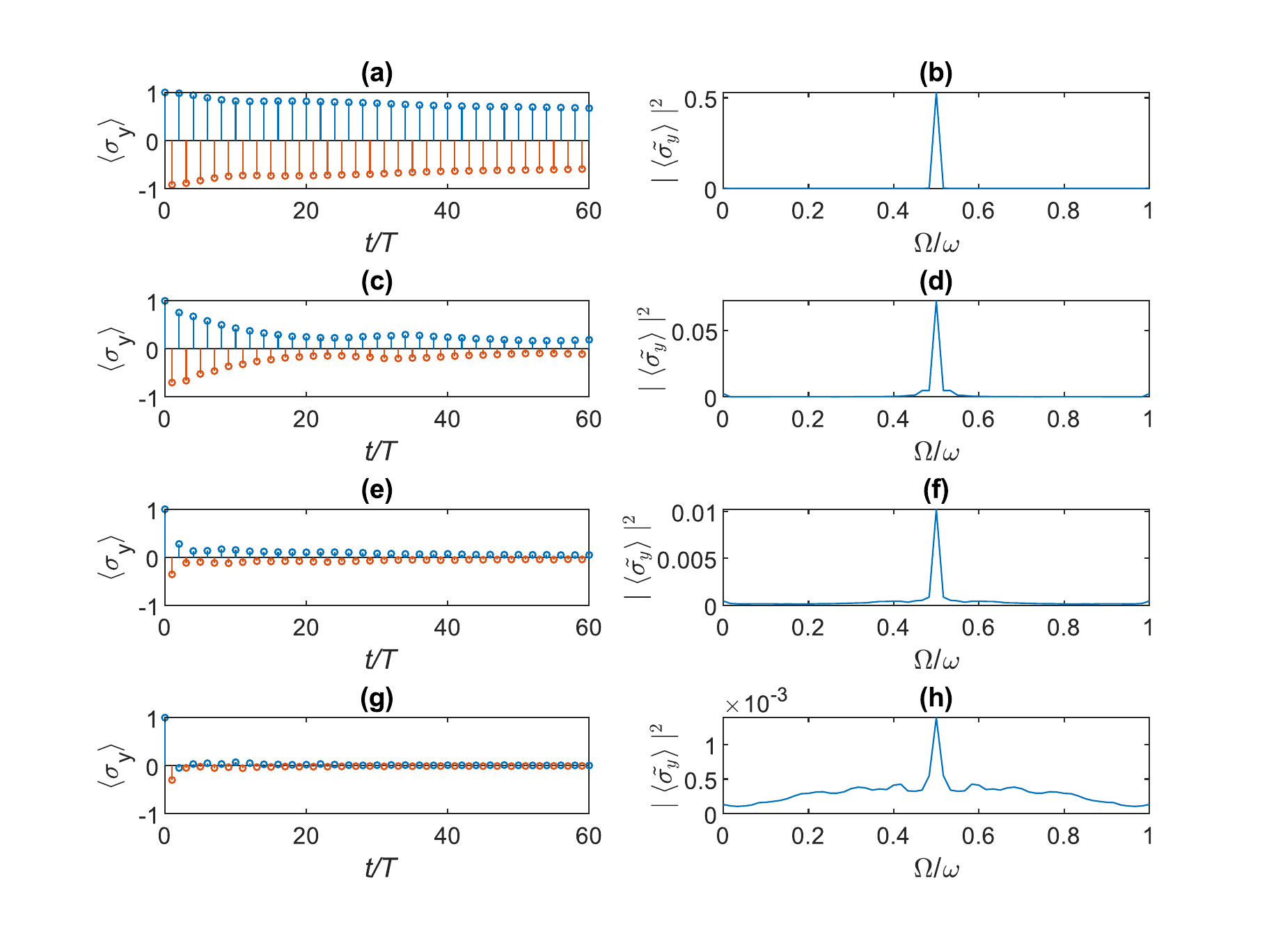}
	\end{center}
	\caption{DMRG results for site number $N=30, M=30, hT/2=\pi/2, \Delta t/T=0.001, \epsilon T=0.05$ and $\lambda T=0.05$. Stroboscopic time evolution of $\langle\sigma_y\rangle$ and its corresponding power spectrum $\mid\langle\tilde{\sigma_y}\rangle\mid^2$ at different interaction strengths (a) \& (b) $JT=0.1$, (c) \& (d) $JT=0.9$, (e) \&  (f) $JT=1.5$, (g) \& (h) $JT=2.0$. In Figs. ((a), (c), (e), (g)), blue (red) dots represent the values of $\langle\sigma_y\rangle$ at even (odd) multiples of the period, whereas vertical lines are to guide the eye at their respective dots.}
	\label{fq3}
\end{figure}

\begin{figure}
	
	\begin{center}
		\includegraphics[scale=1.0]{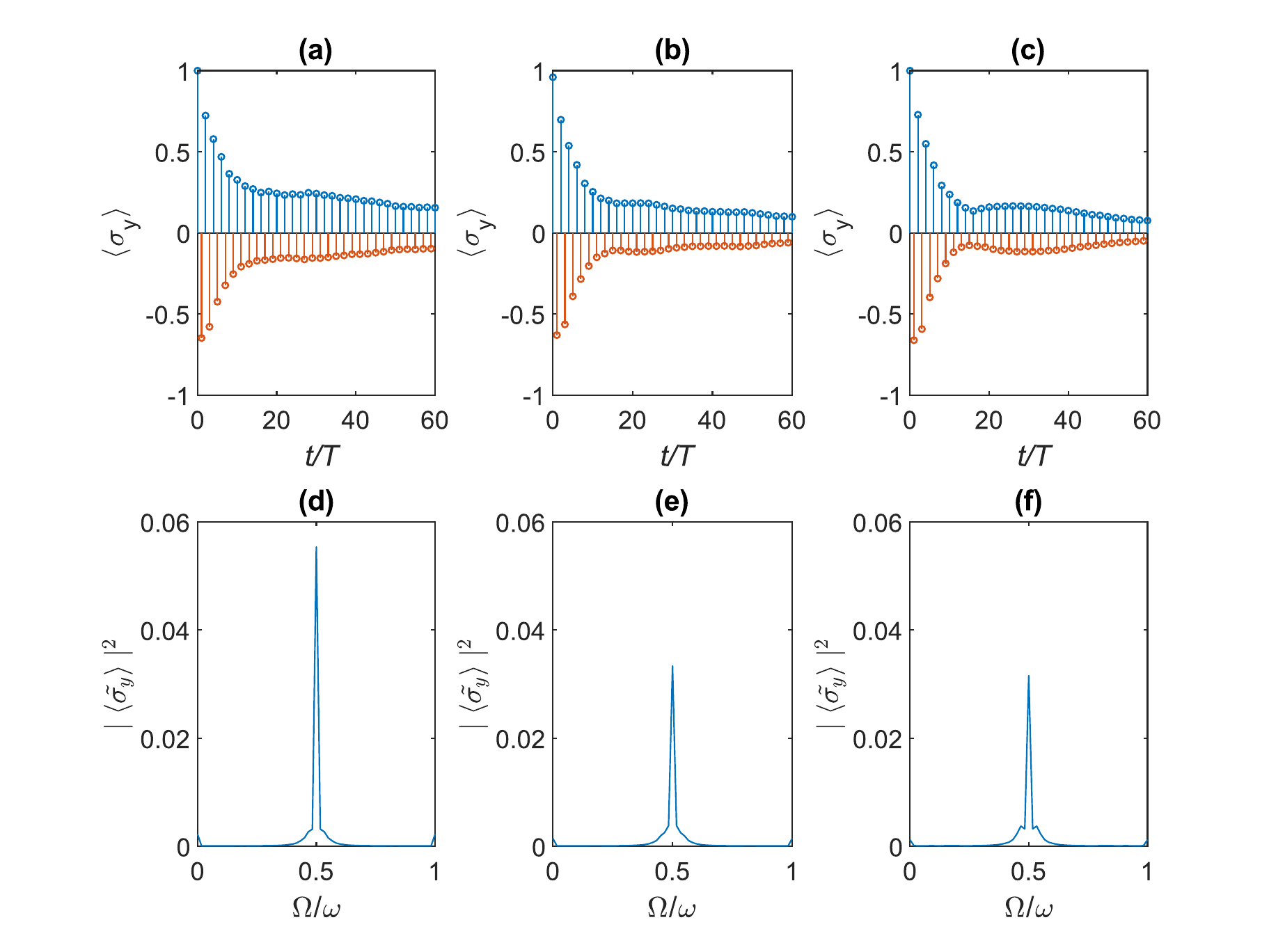}
	\end{center}
	\caption{DMRG results for the maximum order of reduced density matrix $M=30, hT/2=\pi/2, JT=1.0, \Delta t/T=0.001, \epsilon T=0.05$, and $\lambda T=0.05$. Stroboscopic time evolution of $\langle\sigma_y\rangle$ and its corresponding power spectrum $\mid\langle\tilde{\sigma_y}\rangle\mid^2$ at different site numbers (a) \& (d) $N=30$, (b) \& (e) $N=50$, and  (c) \& (f) $N=80$. In Figs. (a)-(c), blue (red) dots represent the values of $\langle\sigma_y\rangle$ at even (odd) multiples of the period, whereas vertical lines are to guide the eye at their respective dots.}
	\label{fq4}
\end{figure}

Finally, we investigate the power spectrum of the total magnetization as the interaction strength is varied. As shown in Fig.~\ref{fq3}, while the peak at $\Omega=\omega/2$ continues to exist at larger interaction strength, it gets smaller as the latter increases. This is consistent with the semiclassical prediction that the system may undergo a transition from the DTC phase to the chaotic regime at sufficiently large interaction strength. On the other hand, while the DTC to chaotic phase transition seems to occur around $JT=3$ in the semiclassical description, it shifts to a smaller value of around $JT\approx 1.5$ in the full quantum setting as evidenced by the extremely small peak in Fig.~\ref{fq3}(e)-(h). This discrepancy arises due to thermalization effect which is not taken into account in the semiclassical description and becomes more significant at larger interaction strength. Indeed, while DTC signatures are observed at small to moderate interaction strength both in semiclassical and full quantum setting, they have different total magnetization dynamics even in the thermodynamic limit where the semiclassical approximation was made. In Fig.~\ref{fq4}, we plot the time evolution of the total magnetization as the system size increases. In the absence of any thermalization, we expect the plot to resemble that of Fig.~\ref{DTCsc}(a) in the thermodynamic limit. Instead, we observe an initial decay in the total magnetization for the first ten periods, followed by a roughly constant magnitude that persists for a long time. Such a structure corresponds to a pre-thermal DTC phase according to Ref.~\cite{the4,pretherm}. Given that the existence of such a pre-thermal DTC phase was only proven for very small interaction strength in Ref.~\cite{the4}, our finding thus verifies that it survives also at moderate interaction strength, as suggested by Fig.~\ref{fq4}.

\section{Concluding remarks}

In this paper, we propose a method for identifying the presence of DTCs in a harmonically driven spin chain. The parameter regime at which DTC phase emerges can be found by invoking a mean-field (semiclassical) approximation to turn the many-body interacting spin system into an effective one-body nonlinear system. By solving the effective one-body Hamilton equations of motion, parameter regime at which DTC phase exists can be identified from the presence and location of the period-doubling islands in its PSOS structure, whose robustness depends on their size. Small to moderate interaction strength is required to observe such period-doubling islands, while larger interaction strength leads to fully chaotic structure. In the presence of general one-body perturbations, two period-doubling islands tend to get closer to each other and eventually merge at large enough perturbations. Still within the mean-field regime, we further define a mean-field magnetization operator and evolve it with time to predict the total magnetization dynamics in the full quantum setting and estimate the largest perturbation strength for which our DTC phase can withstand.

In the full quantum case, we utilize t-DMRG to obtain the time evolution of the total magnetization in the DTC regime predicted by the semiclassical approach, and indeed observe the expected DTC signatures for small perturbation strength. On the other hand, such a DTC phase seems to be destroyed by a smaller interaction strength in the full quantum case due to thermalization effect which is not captured by the mean-field theory. Moreover, in the regime where DTC signatures are observed in the full quantum case, the total magnetization exhibits a pattern typically found in a pre-thermal DTC \cite{the4}. Our results thus suggest that while a semiclassical approximation can be employed to find a regime at which DTC phase exists for a general time-periodic system, the actual DTC signatures in the full quantum limit may be different from those predicted by the semiclassical theory due to quantum effects.  This should be an interesting aspect of quantum chaos in many-body systems.

As a potential future direction, the idea of identifying DTC phases in any interacting time-periodic system from its PSOS structure in the semiclassical limit can also be applied to finding higher-periodicity DTCs (those characterized by observables with $nT$ periodicity where $n>2$), an aspect necessary to realize certain condensed matter phenomena in the time domain \cite{Sacha3,Sacha4,Sacha5,Sacha6,Sacha7,Sacha8}. While it is natural to expect that they may emerge in spin-$(n-1)/2$ systems due to their inherent $Z_n$ symmetry, the possibility of realizing such $nT$-periodic DTCs within the framework of spin-$1/2$ systems is perhaps more surprising and is thus an interesting topic to explore in the future. For example,  the period-quadrupling islands observed in Fig.~\ref{sos}(e) may provide a good starting point along this direction, although presently we are unable to observe the corresponding period-quadrupling total magnetization in the full quantum setting.  There are two reasons behind this. First, such period-quadrupling islands are quite small in size and hence may not be large enough to ``accommodate" a many-body quantum state. Second, these islands in our model only occur at larger interaction strength, where thermalization effect becomes significant. Nevertheless, these are not fundamental obstacles. Rather, we expect that period-quadrupling DTCs can be observed in certain variants of our simple model with period-quadrupling islands in its semiclassical PSOS structure at smaller interaction strength.  We leave this exciting possibility for future work.

\section{Acknowledgment}
P.N. is financially supported by Indonesian Ministry of Research and Higher Education under the contract number 86/UN1/DITLIT/DIT-LIT/LT/2018.
J.G. is supported by the Singapore NRF Grant No. NRF-NRFI2017-
04 (WBS No. R-144-000- 378-281).

\end{document}